# INTERFACE

royalsocietypublishing.org/journal/rsif

## Research

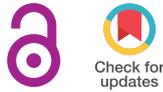

**Cite this article:** Jayles B, Sire C, Kurvers RHJM. 2021 Impact of sharing full versus averaged social information on social influence and estimation accuracy. *J. R. Soc. Interface* **18**: 20210231.
https://doi.org/10.1098/rsif.2021.0231

Received: 19 March 2021
Accepted: 5 July 2021

**Subject Category:**
Life Sciences–Physics interface

**Subject Areas:**
biocomplexity

**Author for correspondence:**
Bertrand Jayles
e-mail: bertrand.jayles@ntu.edu.sg

Electronic supplementary material is available online at https://doi.org/10.6084/m9.figshare. c.5513345.

**THE ROYAL SOCIETY** PUBLISHING
# Impact of sharing full versus averaged social information on social influence and estimation accuracy

Bertrand Jayles[1,2], Clément Sire[3] and Ralf H. J. M. Kurvers[1]

[1]Center for Adaptive Rationality, Max Planck Institute for Human Development, Lentzeallee 94, 14195 Berlin, Germany
[2]Institute of Catastrophe Risk Management, Nanyang Technological University, Block N1, Level B1b, Nanyang Avenue 50, Singapore 639798, Republic of Singapore
[3]Laboratoire de Physique Théorique, Centre National de la Recherche Scientifique (CNRS), Université de Toulouse—Paul Sabatier (UPS), Toulouse, France

BJ, 0000-0002-4550-7919; CS, 0000-0003-4089-4013; RHJMK, 0000-0002-3460-0392

The recent developments of social networks and recommender systems have dramatically increased the amount of social information shared in human communities, challenging the human ability to process it. As a result, sharing aggregated forms of social information is becoming increasingly popular. However, it is unknown whether sharing aggregated information improves people's judgments more than sharing the full available information. Here, we compare the performance of groups in estimation tasks when social information is fully shared versus when it is first averaged and then shared. We find that improvements in estimation accuracy are comparable in both cases. However, our results reveal important differences in subjects' behaviour: (i) subjects follow the social information more when receiving an average than when receiving all estimates, and this effect increases with the number of estimates underlying the average; (ii) subjects follow the social information more when it is higher than their personal estimate than when it is lower. This effect is stronger when receiving all estimates than when receiving an average. We introduce a model that sheds light on these effects, and confirms their importance for explaining improvements in estimation accuracy in all treatments.

## 1. Introduction

Social information is a crucial component of human and animal decision-making [1,2]. Most of people's everyday choices, whether picking a movie or a restaurant, finding the best school for their children or gathering information before voting in an election, are influenced by the experiences and opinions of others [3]. From a broader perspective, social learning strategies, which consist in exploiting social information selectively, continue to play a central role in the emergence and evolution of cultures and their startling diversity [4,5]. Understanding the impact of social information on judgment and decision-making is thus crucial for comprehending individual and collective behaviour in human and animal groups [3,6].

Information technology has altered how people relate to information and how individuals interact with and influence each other. People are more connected to each other than ever before: social networks, blogs and websites, and the massive diffusion of smartphones have made information and virtual others instantaneously available, anywhere and at any time [7]. Moreover, online recommender systems and social networks have considerably extended people's exposure to others' opinions and recommendations [8–10]. For instance, when selecting a restaurant, a travel destination or a hotel, the first thing one often does is look at others' ratings and reviews. This permanent exchange of social information, generally mediated by digital interfaces, is likely to amplify in the coming years, with new generations being born and raised with smartphones and the Internet. This brings about new challenges, such as how to process so much information and make efficient judgments, especially given people's limited time and cognitive resources [11–13]. One issue of particular importance is how to best exchange social information in human groups in a way that improves individual and collective judgments: while aggregated social information is easy to process, providing individuals access to all the available information may give them more ground to make better judgments. From a pure information theoretic perspective, having more or better quality information should indeed lead to better decisions, although more complex information may also challenge human cognitive limits.

Here, we address this important issue through the prism of estimation tasks, a highly suitable paradigm for quantitative studies on social influenceability [14–21]. We performed experiments in which subjects in groups of 12 members were asked to estimate a series of quantities both before and after receiving social information. Social information consisted of a varying number of estimates $\tau$ from other group members ($\tau$ = 1, 3, 5, 7, 9 or 11). Three treatments were tested: subjects either received (i) all $\tau$ estimates in ascending order ('Sorted' treatment), (ii) all $\tau$ estimates in a random order ('Unsorted' treatment), or (iii) the geometric mean of the $\tau$ estimates ('Aggregated' treatment). Crucially, in the Aggregated treatment, and contrary to previous studies [20,21], subjects were made aware of the number of estimates used to calculate the geometric mean. Previous studies have analysed the patterns of social influenceability and the conditions under which social influence can improve estimation accuracy [14–18,20–24]. However, to the best of our knowledge, a direct and systematic comparison of the effects of averaged versus full social information on collective judgments in estimation tasks, with a focus on the number of estimates shared, has been lacking.

Our results show that subjects are more sensitive to social influence when receiving the geometric mean of the shared estimates than when receiving all shared estimates. Moreover, the sensitivity to social influence depends on the distance between their personal estimate and the geometric mean of the shared estimates. We then build and calibrate an agent-based model exploiting these findings, and use it to analyse the impact of the number of shared estimates $\tau$ on social influenceability and estimation accuracy. We find that, in the Sorted and Aggregated treatments, sensitivity to social influence increases with $\tau$ and then saturates. We also show that improvements in individual accuracy increase with $\tau$ before saturating, and are comparable in all treatments.

## 2. Experimental design

A total of 216 subjects (138 females, 70 males, 8 unreported; mean age ± s.d.: 26 ± 4, 17 unreported), distributed over 18 groups of 12 individuals, took part in the experiment. All individuals signed an informed consent form prior to participating. The experiment was approved by the Institutional Review Board of the Max Planck Institute for Human Development (ARC 2018/08). Each individual was confronted with 42 estimation questions on a tactile tablet (see electronic supplementary material, figure S1). Questions ranged from estimating the number of marbles in a jar, to the population of Tokyo, to the number of stars in the Milky Way (see electronic supplementary material for a list of all questions). Each question was asked twice: first, subjects provided their personal estimate $E_P$. Next, they were given the estimate(s) of one or several group member(s) as social information and were then asked to provide a second estimate $E_s$ (figure 1). The social information never contained a participant's own estimate.

Social information was displayed to the subjects in three ways: (i) the 'Sorted treatment', where $\tau$ estimates ($\tau$ = 1, 3, 5, 7, 9, 11) were displayed in ascending order; (ii) the 'Unsorted treatment', where the $\tau$ estimates were displayed in random order; and (iii) the 'Aggregated treatment', where only the geometric mean of the $\tau$ estimates was displayed. Note that in the Sorted and Unsorted treatments, the $\tau$ pieces of social information were shown simultaneously. In all treatments, the shared estimates were selected randomly, and in the Aggregated treatment subjects were informed about the number of estimates $\tau$ used to compute the geometric mean. Participants in each group only experienced one of the three display treatments (i.e. a between-subject design). This was done to avoid the 'leakage' of strategies and/or information across treatments. For instance, being exposed to the dispersion of estimates in the Sorted or Unsorted treatment may impact a person's subsequent decisions on the integration of social information in the Aggregated treatment. The number $\tau$ of estimates shared did, however, vary within groups. Overall, there were 18 conditions (3 treatments times 6 values of $\tau$), and 504 estimates per condition (42 questions times 12 subjects), both before and after social information sharing.

The 42 questions were randomly assigned to seven blocks of six questions. Across groups, the order of the blocks and the questions within a block were randomized. A block always contained each number of estimates to be shared (1, 3, 5, 7, 9 and 11) once. All subjects thus experienced each level of $\tau$ the same amount of times. The randomization was constrained so that across all of the 18 groups, each unique question was asked once at each unique combination of display (three levels) and number of estimates shared (six levels). All tablets were controlled by a central server, and participants could only proceed to the next question once all individuals had provided their second estimate. A 30-s countdown timer was shown on the screen to motivate subjects to answer within this time window, although they were allowed to take more time. Subjects received a flat fee of €15 for participation and a bonus payment of €1 to €5 depending on their performance (see electronic supplementary material for detailed payment information).

## 3. Results

### 3.1. Distribution of estimates

Because of the human logarithmic internal representation of numbers [25], it is more appropriate to consider logarithms of estimates than the estimates themselves in estimation tasks [20,26]. This logarithmic representation also motivated us in presenting the geometric—rather than the arithmetic—mean of $\tau$ estimates to the subjects in the Aggregated treatment. Moreover, to make estimates of different quantities comparable, it is necessary to normalize them by the true value of their respective quantities [20,21]. We therefore use the quantity $X = \log(E/T)$ as our variable of interest, where $E$ is the actual estimate of a quantity and $T$ the corresponding true value. $X$ represents the deviation from the truth in orders of magnitude. For simplicity, we will, from now on, refer to $X$ as 'estimates', with $X_p$ being personal estimates and $X_s$ being second estimates (i.e. after receiving the social information). Figure 2 shows the distributions of $X_p$ (closed dots) and $X_s$ (open dots) in each treatment and number of shared estimates $\tau$. Distributions of personal estimates $X_p$ for individual questions are shown in electronic supplementary material, figure S2.

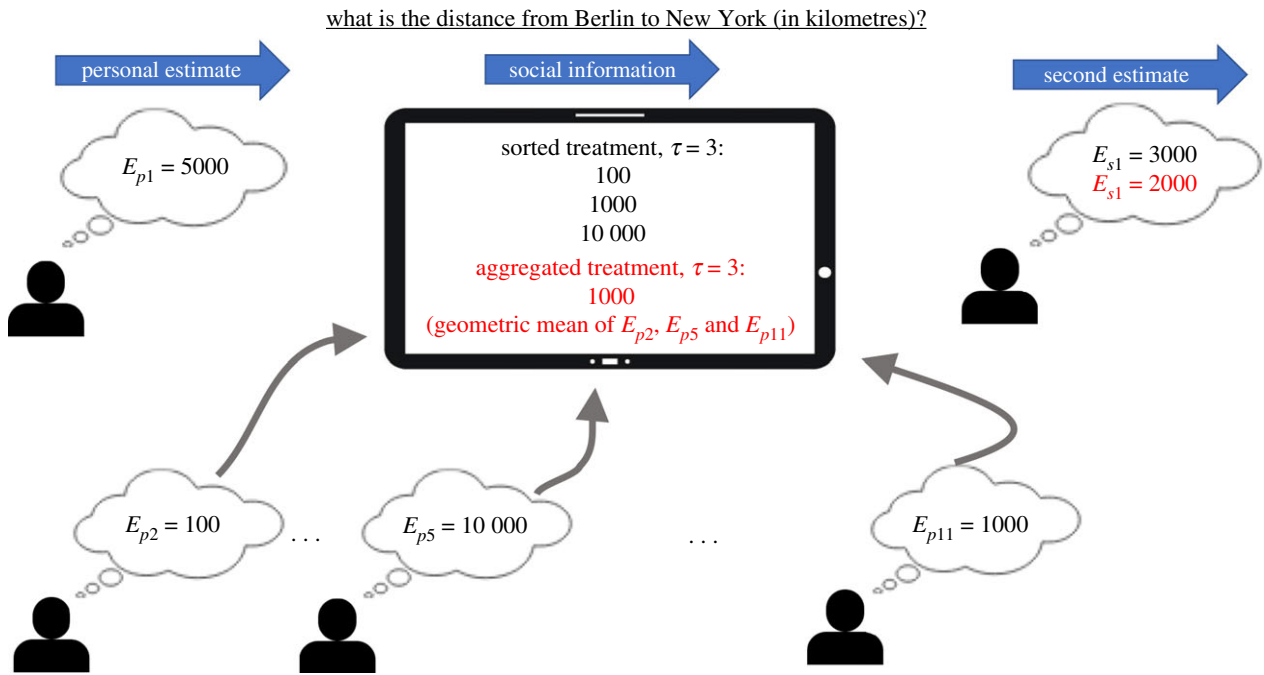

**Figure 1.** Overview of the experimental procedure. First, all 12 participants independently provide their personal estimate to a given question ($E_{p1} \ldots E_{p12}$). Then, social information is exchanged. In this specific example, participant 1's personal estimate is $E_{p1} = 5000$, and receives as social information the personal estimates of participants 2, 5 and 11 ($E_{p2}$, $E_{p5}$ and $E_{p11}$). In the Sorted treatment (black), participant 1 is shown these $\tau = 3$ estimates in increasing order (and in the Unsorted treatment (not shown) in random order). In the Aggregated treatment (red), participant 1 is shown the geometric mean of the $\tau = 3$ estimates. Finally, each participant provides a second estimate (as shown for participant 1, $E_{s1}$).

We find, in agreement with other studies [20,26], that the distributions of estimates peak close to $X = 0$ (i.e. near $E = T$, that is the true value; see below), a phenomenon known as the 'wisdom of crowds' [27], and that these distributions narrow after social information sharing (i.e. lower dispersion of the second estimates driven by social influence; see electronic supplementary material, figure S3). Previous studies showed that the distribution of estimates for a given quantity is often close to a Laplace distribution [26,28], of probability density function $P(X) = (1/(2w))\exp(-|X-c|/w)$.

The solid lines in figure 2 simulate distributions of personal estimates $X_p$, where the $X_p$ are drawn, for each question, from Laplace distributions, the centre $c$ and width $w$ of which are, respectively, the median and dispersion (average absolute deviation from the median) of the experimental personal estimates for that question. Electronic supplementary material, figure S4, shows the distribution of $X_p$ for all conditions combined, supporting the Laplace distributions assumption. The dashed lines in figure 2 are predictions of the model—presented below—for the distributions of second estimates $X_s$. Note that since for the questions asked in this experiment, actual estimates $E_{p,s}$ cannot be lower than 1, the $X_{p,s}$ are simulated with the restriction that $X_{p,s} > -\log(T)$, which imposes a sharper decay on the left side of the distribution.

### 3.2. Sensitivity to social influence

We define, in all treatments, the *sensitivity to social influence* $S$ of a subject answering a specific question, as the (barycentric) weight they give to the average $M = \log(G)$ of the social information received for this question, where $G$ is the geometric mean of the shared estimates (see Statistical methods). With this definition, the second estimate is hence given by $X_s = (1-S)X_p + SM$. Note that contrary to the Aggregated treatment where $M$ is provided to the subjects, the subjects in the Sorted and Unsorted treatments do not have a direct access to $M$ (except for $\tau = 1$). However, the knowledge of $X_s$ and $X_p$ allows us to uniquely define $S$ even in these cases. In fact, in the Sorted and Unsorted treatments, the participants tend to focus their attention on the central tendency of the social information when receiving multiple estimates. This is supported by previous findings [22,29,30], and is consistent with the use of heuristic strategies under time and cognitive constraints [11–13].

#### 3.2.1. Distribution of *S*

Figure 3 shows the distributions of $S$ in all treatments and number of shared estimates $\tau$. The experimental distributions (solid lines) consist of a peak at $S = 0$ (i.e. subjects keeping their initial estimate) and a part that we approximate in the model as a Gaussian distribution, consistent with previous studies [20,26]. We formalize this by assuming that subjects keep their initial estimate ($S = 0$) with probability $P_0$, or draw an $S$ from a (close to) Gaussian distribution of mean $m_g$ and standard deviation $\sigma_g$ with probability $P_g = 1 - P_0$ (index g referring to 'Gaussian'). This imposes

$$\langle S \rangle = P_0 \times 0 + P_g m_g = P_g m_g, \quad \text{i.e. } P_g = \frac{\langle S \rangle}{m_g}, \quad (3.1)$$

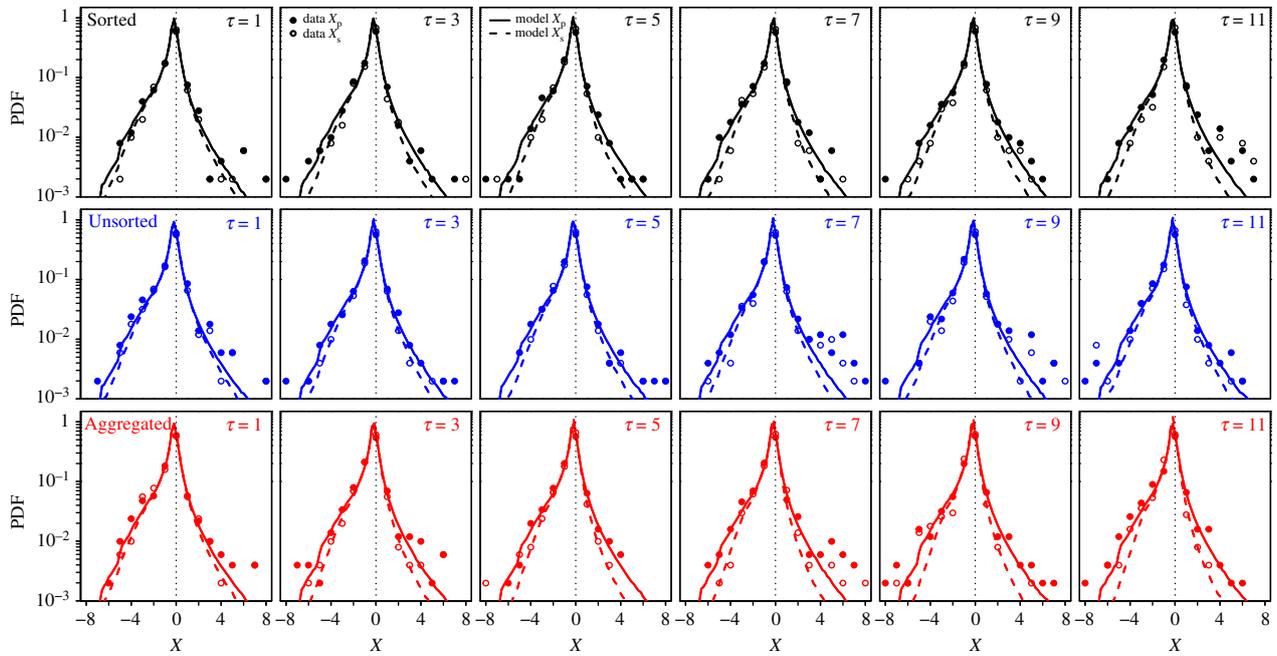

**Figure 2.** Probability density function (PDF) of personal estimates $X_p$ (closed dots) and second estimates $X_s$ (open dots) in the Sorted (black), Unsorted (blue) and Aggregated (red) treatments, for all values of $\tau$. Lines are model simulations, with solid lines representing distributions of $X_p$ and dashed lines distributions of $X_s$.

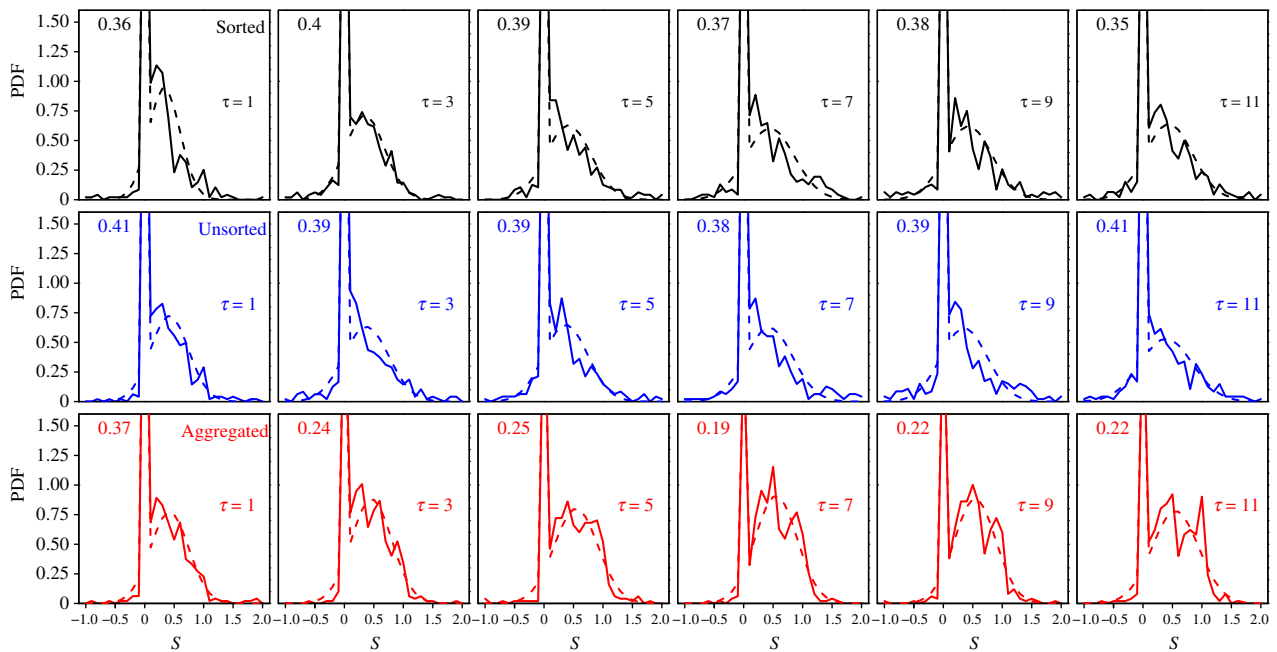

**Figure 3.** Probability density function (PDF) of sensitivities to social influence $S$ in the Sorted (black), Unsorted (blue) and Aggregated (red) treatments, for all values of $\tau$. Solid lines are experimental data, and dashed lines fits using hurdle models (see Statistical methods). The weight (i.e. percentage of occurrences) $P_0$ at $S = 0$ is shown at the topleft part of each panel.

with $\langle S \rangle$ denoting the mean of $S$ for a given treatment and value of $\tau$.

We fit a hurdle model (i.e. a Dirac peak for 0 values and a Gaussian distribution for non-zero values; see Statistical methods) to the experimental distributions to estimate the values of $P_g$, $m_g$ and $\sigma_g$ per condition. As a consistency check, electronic supplementary material, figure S5a, shows that the fit-predicted values of the density at $S = 0$ match the experimental values well.

Note that in the above-mentioned studies, where subjects received the average of a few previous estimates as social information, a small peak at $S = 1$ (i.e. subjects adopting the social information) was observed (about 4% of the data). In the Sorted and Unsorted treatments of our study, an $S$ of exactly 1 is unlikely when $\tau > 1$ because subjects cannot easily calculate the exact geometric mean of several estimates. The percentage of $S = 1$ was indeed 0 in these conditions, except at $\tau = 3$ in the Sorted treatment (2 instances, 0.4% of the data). In the Aggregated treatment, $S = 1$ is in principle possible when $\tau > 1$, but also in these conditions there were no instances of $S = 1$. Only at $\tau = 1$ did we observe this with a probability of at most 2%. We, therefore, decided not to include a separate peak at $S = 1$ in the model presented below, but to absorb the very few instances of $S = 1$ in the Gaussian part of the distribution.

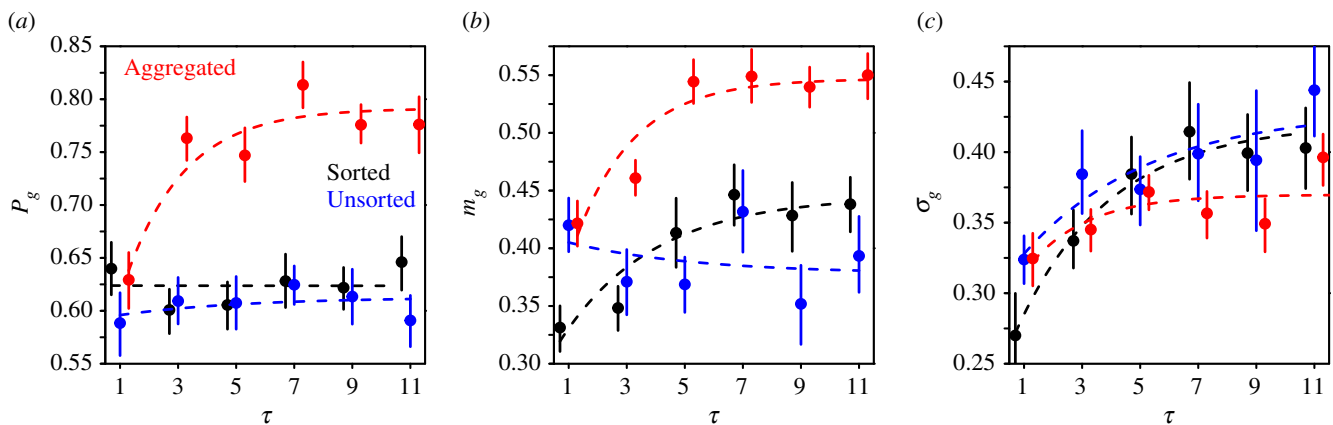

**Figure 4.** Dependence of (a) $P_g$, (b) $m_g$ and (c) $\sigma_g$ on the number of estimates shared $\tau$, in the Sorted (black), Unsorted (blue) and Aggregated (red) treatments. Dots represent the data, and dashed lines are fits using exponential saturation functions (see Statistical methods). Error bars are computed using a bootstrap procedure described in the Statistical methods, and roughly represent 1 s.e.

Note that some of the values of the distribution at $S = 1$ in figure 3 appear different from 0, especially in the Aggregated treatment. This is, however, a result of the binning (bin size of 0.1, meaning that values of $S$ between 0.95 and 1.05 were included for $S = 1$). Figure 4 shows how the fitted $P_g$, $m_g$ and $\sigma_g$ vary with $\tau$ and treatment.

These quantities are found to either saturate or remain constant (which amounts to an immediate saturation). A saturation is expected at large values of $\tau$, in particular since the probability $P_g$ must remain between 0 and 1. In the Aggregated treatment, for instance, it would be surprising if subjects' behaviour would change dramatically when receiving the average of 110 estimates compared to the average of 100 estimates. Likewise, in the Sorted and Unsorted treatments, we expect a saturation due to cognitive limitations. There must be some limit in the number of estimates that the subjects can properly process, beyond which they would start using heuristics such as sampling only a few estimates around the central value (geometric mean) of the social information. Beyond this point, the number of shared estimates should also not matter anymore. We thus fit the data with an exponential saturation (the simplest saturation form) and assume that the three quantities saturate at the same rate $\varepsilon$ (see details in Statistical methods). Theoretically, one could fit a separate saturation rate for each parameter but this would make the model unnecessarily complicated. Indeed, this saturation assumption yields a very good fit with the experimental results.

Interestingly, the saturation happens at relatively low values of $\tau$ ($\tau = 3$–5). In the Sorted treatment, $P_g$ is independent of $\tau$, while $m_g$ and $\sigma_g$ increase with $\tau$. In the Unsorted treatment, $P_g$ and $m_g$ barely depend on $\tau$, while $\sigma_g$ increases with $\tau$, although less strongly than in the Sorted treatment. In the Aggregated treatment, $P_g$ and $m_g$ strongly increase with $\tau$, with substantially higher values than in both other treatments. Subjects thus compromise substantially more with the social information when receiving an average than when receiving multiple estimates, and increasingly so when the average is based on more estimates. $\sigma_g$ is also found to increase with $\tau$ in the Aggregated treatment, but at a milder rate than in both other treatments, and with lower values on average. Note that at $\tau = 1$, one may expect all treatments to have the same results, since in all three cases subjects receive a single random estimate. However, at $\tau = 1$, $m_g$ and $\sigma_g$ were significantly lower in the Sorted treatment than in both other treatments. In each treatment, subjects experienced each level of $\tau$ seven times, which may have affected their behaviour at $\tau = 1$ in different ways. It is possible that repeatedly receiving multiple sorted estimates (in the Sorted treatment) negatively affected subjects' trust in the social information when experiencing $\tau = 1$, making them compromise less (i.e. lower values of $m_g$).

### 3.2.2. Impact of $D = M − X_p$ on sensitivity to social influence

Previous studies [20,26] have shown that the average weight given to the social information $\langle S \rangle$ grows linearly with $|D|$, where $D = M − X_p$ is the distance between the personal estimate $X_p$ and the social information $M$:

$$\langle S \rangle(D) = \alpha + \beta |D|. \quad (3.2)$$

We call this the *distance effect*. In these studies, subjects were presented the average $M$ of an *unknown* number of estimates from other group members. At variance, in our experiment, subjects were presented all pieces of social information in the Sorted and Unsorted treatments, and were aware of the number of estimates underlying $M$ in the Aggregated treatment. In both cases, we thus expect $\alpha$ and $\beta$ to depend on $\tau$:

$$\langle S \rangle(\tau, D) = \alpha(\tau) + \beta(\tau) |D|. \quad (3.3)$$

Figure 5 shows the distance effect in all treatments and number of shared estimates $\tau$. In the Sorted treatment, the bottom of the cusp is, however, not at $D = 0$, but at $D = D_0 < 0$, suggesting that subjects follow the social information the least when its geometric mean is slightly lower than their personal estimate. This effect is substantially less marked in the Aggregated treatment, and entirely absent in the Unsorted treatment.

The fitting parameters are $D_0$, $\alpha$ (the bottom of the cusp), and the (possibly different) slopes $\beta_−$ for $D < D_0$ and $\beta_+$ for $D > D_0$. We estimate them by minimizing squared errors with the following function:

$$\langle S \rangle(\tau, D) = \alpha(\tau) + \beta_\pm(\tau) |D − D_0(\tau)|, \quad (3.4)$$

where $\beta_\pm = \beta_−$ when $D < D_0$ and $\beta_\pm = \beta_+$ when $D > D_0$ (see Statistical methods for details of the fitting procedure).

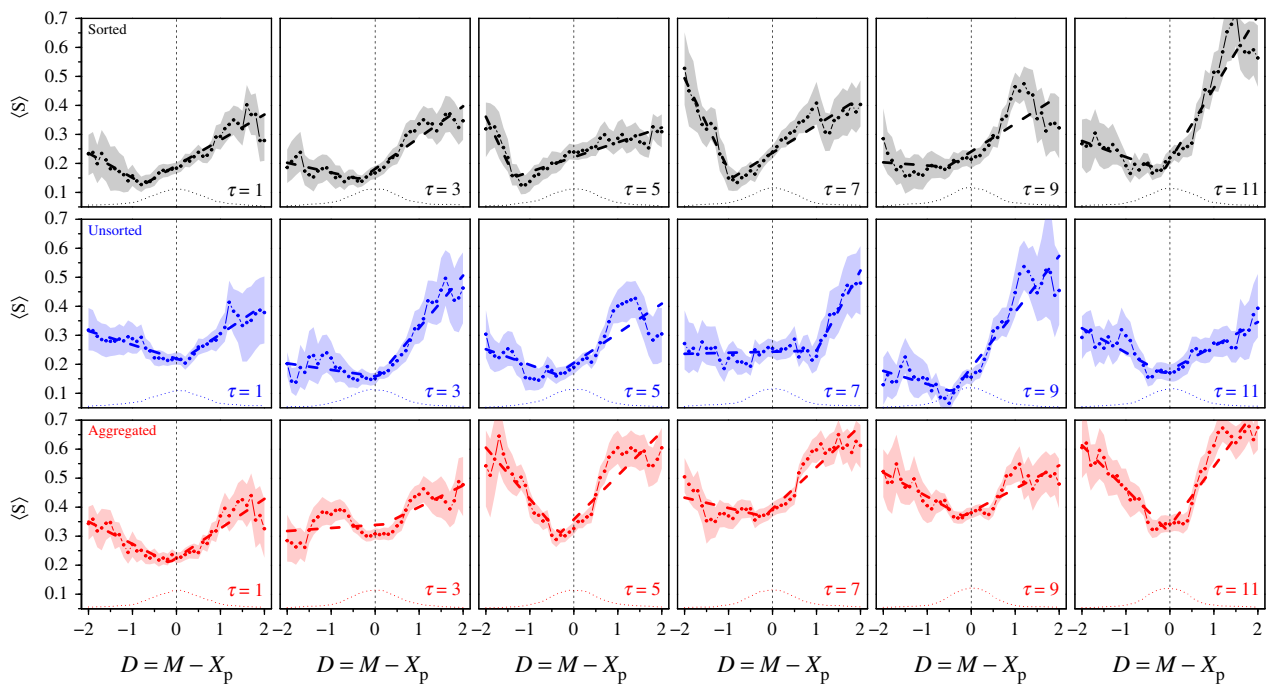

**Figure 5.** Average sensitivity to social influence $\langle S \rangle$ against the distance $D = M - X_p$ between personal estimate $X_p$ and social information $M$, in the Sorted (black), Unsorted (blue) and Aggregated (red) treatments, for all values of $\tau$. Dots are the data, and shaded areas roughly represent 1 s.e., and were computed using a bootstrap procedure described in the Statistical methods. Dashed lines are least-squares fits (see Statistical methods) and dotted bottom curves show the density of data at a given $D$ in arbitrary units.

Figure 6 shows $D_0$, $\alpha$, $\beta_-$ and $\beta_+$ against $\tau$ in all treatments. Since we do not observe any clear dependence of $D_0$, $\beta_-$ and $\beta_+$ on $\tau$, we will consider them constant with $\tau$ in the model presented below, and equal to their average value over all values of $\tau$. Note that in all treatments, $\beta_-$ and $\beta_+$ are of the same order as the slope obtained in [20,26].

The parameter $\alpha$, however, underlies the exponential saturation of $P_g$ (figure 4a). For the sake of consistency, $\alpha$ therefore needs to be fitted with the same saturation form as used in figure 4, and with the same rate $\varepsilon$. Consistent with figure 4, $\alpha$ is found to slowly increase with $\tau$ in the Sorted treatment, while it increases rapidly in the Aggregated treatment, with substantially higher values. In the Unsorted treatment, $\alpha$ is found to slightly decrease with $\tau$, with similar values as in the Sorted treatment.

We will therefore use, in our model, the following general equation for $\langle S \rangle$:

$$\langle S \rangle (\tau, D) = \alpha_\infty - (1-\varepsilon)^{(\tau-1)} (\alpha_\infty - \alpha_1) + \beta_\pm |D - D_0|, \quad (3.5)$$

where $\alpha_\infty$ is the saturation value of $\alpha$ and $\alpha_1$ its value at $\tau = 1$.

### 3.3. Model of social information integration
To analyse the effects of the number of shared estimates $\tau$ and the display treatment on sensitivity to social influence and estimation accuracy, we introduce an agent-based model.

The model is inspired by a model developed in [20,26], where subjects received as social information the average of an *unknown* number of estimates from other group members. For a given display treatment, value of $\tau$, and quantity to estimate, the model proceeds in three steps:

1. *Personal estimates*: first, personal estimates $X_p$ are drawn, for all agents in groups of size $N$, from a Laplace distribution, the centre and width of which are, respectively, the median $m_p$ and dispersion $\sigma_p = \langle |X_p - m_p| \rangle$ of the experimental personal estimates of the quantity.

2. *Social information*: next, each agent receives as social information the average $M$ of $\tau$ personal estimates from randomly selected other group members. Each agent keeps its personal estimate ($S = 0$) with probability $P_0$ or draws an $S$ from a Gaussian distribution of mean $m_g$ and standard deviation $\sigma_g$ with probability $P_g$:

$$P_g(\tau, D) = \frac{\langle S \rangle (\tau, D)}{m_g(\tau)} = \frac{(\alpha(\tau) + \beta_\pm |D - D_0|)}{m_g(\tau)}. \quad (3.6)$$

$P_0$ is given by $P_0 = 1 - P_g$.

3. *Second estimates*: finally, once $P_g$ and $P_0$ are determined for each agent, and an $S$ is drawn accordingly, each agent's second estimate $X_s$ is calculated as the weighted average of its personal estimate $X_p$ and the mean $M$ of the social information (by definition of $S$):

$$X_s = (1-S) X_p + S M. \quad (3.7)$$

The parameter values are based on the fits in figures 4b,c and 6, and are provided in table 1. Each run of the model mimicked our experiment. The results of the model simulations shown in the figures are averages over 10 000 runs. We next test the model on measures that were not used in its design and calibration.

### 3.4. Impact of $\tau$ on sensitivity to social influence
Figure 7a shows the average sensitivity to social influence $\langle S \rangle$ as a function of $\tau$ in all treatments. Again, the symbol $\langle \bullet \rangle$ denotes an average over all the relevant data.

As a consistency check of the relation $\langle S \rangle = P_g m_g$ (equation (3.1)), figure 7b shows $P_g m_g$ against $\tau$, where $P_g$ and $m_g$ take the values estimated in figure 4a,b. The fits in this graph were obtained by multiplying the analytical expressions of $P_g$ and $m_g$ (see equation (5.3) in the Statistical methods), using the parameter values estimated in figure 4a,b.

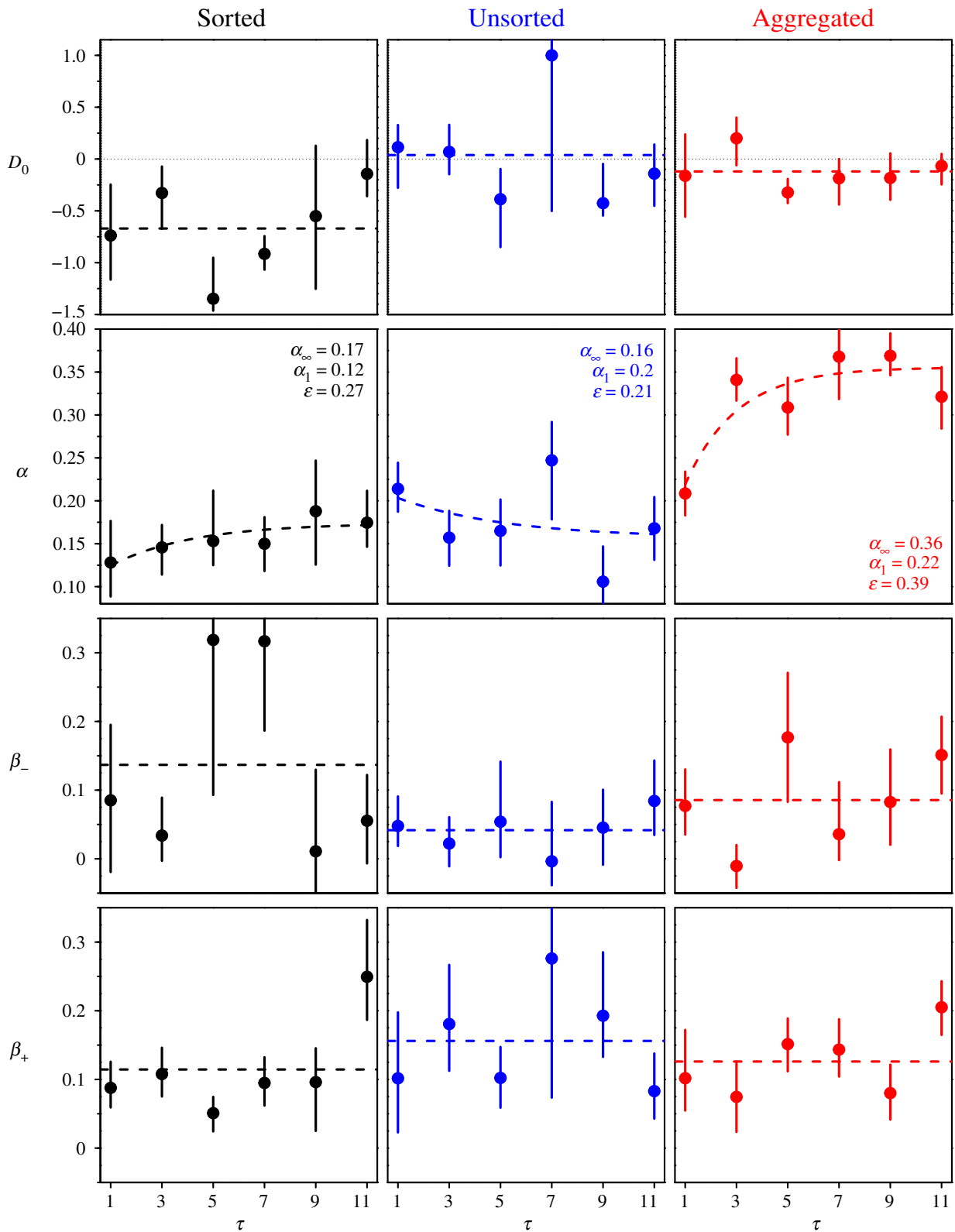

**Figure 6.** Dependence of $D_0$, $\alpha$, $\beta_-$ and $\beta_+$ on $\tau$ in the Sorted (black), Unsorted (blue) and Aggregated (red) treatments. Error bars were computed using a bootstrap procedure described in the Statistical methods, and roughly represent 1 s.e. Dashed lines show the values used in the model presented below: $D_0$, $\beta_-$ and $\beta_+$ are assumed constant, and taken as the mean over all values of $\tau$. $\alpha$, however, which drives the exponential saturation of $P_g$ (figure 4a), is fitted with the same saturation form and at the same rate $\varepsilon$ as in figure 4.

The fits are in good agreement with the data, and show that $\langle S \rangle$ increases slowly with $\tau$ in the Sorted treatment, and sharply in the Aggregated treatment. Subjects thus give substantially more weight to the social information when receiving averages than when receiving all pieces of social information, and increasingly so as the average is based on more estimates. In both treatments and as expected, $\langle S \rangle$ saturates at intermediate values of $\tau$. In the Unsorted treatment, a very small decrease is observed before saturation. These patterns are consistent with the patterns of $P_g$ (figure 4a), $m_g$ (figure 4b) and $\alpha$ (figure 6 second row) against $\tau$. Note that these fits are not inputs of the model, which predicts a slightly milder increase with $\tau$ in the Sorted and Aggregated treatments, and a slightly faster decrease with $\tau$ in the Unsorted treatment (figure 7a). Note that consistency with equation (3.5) also requires that $\langle S \rangle_\infty = \alpha_\infty + \langle \beta_\pm | D - D_0 | \rangle_D$. Electronic supplementary material, figure S5b, compares the saturation value of $\langle S \rangle_\infty$ fitted in figure 7b against the quantity $\alpha_\infty + \langle \beta_\pm | D - D_0 | \rangle_D$ in all treatments, where $\beta_-$, $\beta_+$ and $D_0$ have been given the average values found in figure 6, and $\alpha_\infty$ the saturation value fitted in the same figure. The three points are, as expected, close to the diagonal line.

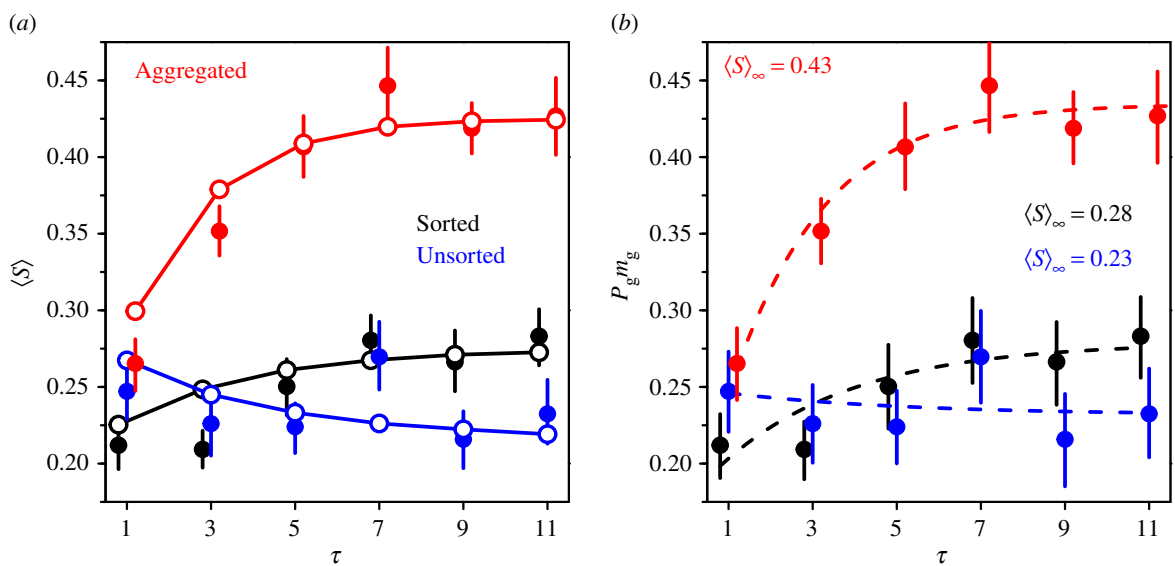

**Figure 7.** Dependence of (a) the average sensitivity to social influence $\langle S \rangle$ and (b) $P_g m_g$ on the number of shared estimates $\tau$, in the Sorted (black), Unsorted (blue) and Aggregated (red) treatments. Solid lines and empty circles in (a) are model simulations and dashed lines in (b) are fits produced by multiplying the analytical expression of $P_g$ and $m_g$ (equation (5.3)), using the parameter values estimated in figure 4a,b. Error bars in (a) were computed using a bootstrap procedure described in the Statistical methods, and roughly represent 1 s.e. Error bars in (b) were computed by multiplying $P_g = P_{g_{exp}} \pm \Delta_{P_g}$ and $m_g = m_{g_{exp}} \pm \Delta_{m_g}$, where $P_{g_{exp}}$ and $m_{g_{exp}}$ are the values estimated in figure 4a,b, and $\Delta_{P_g}$ and $\Delta_{m_g}$ the corresponding errors. The fitted saturation value $\langle S \rangle_\infty$ is shown in each treatment in (b).

**Table 1.** Parameter values used in the model for each treatment.

| parameter | description | Sorted | Unsorted | Aggregated |
|---|---|---|---|---|
| $\varepsilon$ | saturation rate | 0.27 | 0.21 | 0.39 |
| $m_{g1}$ | value of $m_g$ at $\tau = 1$ | 0.32 | 0.41 | 0.41 |
| $m_{g\infty}$ | saturation value of $m_g$ | 0.45 | 0.38 | 0.55 |
| $\sigma_{g1}$ | value of $\sigma_g$ at $\tau = 1$ | 0.27 | 0.33 | 0.32 |
| $\sigma_{g\infty}$ | saturation value of $\sigma_g$ | 0.42 | 0.43 | 0.37 |
| $\alpha_1$ | value of $\alpha$ at $\tau = 1$ | 0.12 | 0.20 | 0.22 |
| $\alpha_\infty$ | saturation value of $\alpha$ | 0.17 | 0.16 | 0.36 |
| $D_0$ | bottom of the cusp relationship | −0.67 | 0.04 | −0.12 |
| $\beta_-$ | left slope of the cusp relationship | 0.14 | 0.04 | 0.09 |
| $\beta_+$ | right slope of the cusp relationship | 0.11 | 0.16 | 0.13 |

Figure 8 shows that subjects follow the social information more (i.e. $\langle S \rangle$ is higher) when the average social information is higher than their personal estimate $(D > 0)$ than when it is lower $(D < 0)$. Though the strength of the effect slightly differs between treatments, it is present in all conditions. This result is a combined effect of the above findings that $D_0 < 0$ and $\beta_+ > \beta_-$. In the next section, we will show that this effect has an impact on estimation accuracy after social information sharing.

### 3.5. Impact of $\tau$ on estimation accuracy

Following [20], we define (i) *collective accuracy* as $|\text{Median}_{i,q}(X_{i,q})|$, where $i$ runs over individuals and $q$ over quantities, and (ii) *individual accuracy* as $\text{Median}_{i,q}(|(X_{i,q})|)$. Values closer to 0 (the log-normalized transformation of the truth is 0) indicate better accuracy for both measures. Collective accuracy measures how close the median estimate of all group members is to the truth, and individual accuracy measures how close, on average, individual estimates are to the truth. Collective and individual accuracies are different, but related, aspects of accuracy. An improvement in collective accuracy amounts to a shift of the median estimate towards the truth, which is perforce accompanied by an improvement in individual accuracy, as individual estimates also get, on average, closer to the truth. However, there can be an improvement in individual accuracy without an improvement in collective accuracy if estimates converge after social

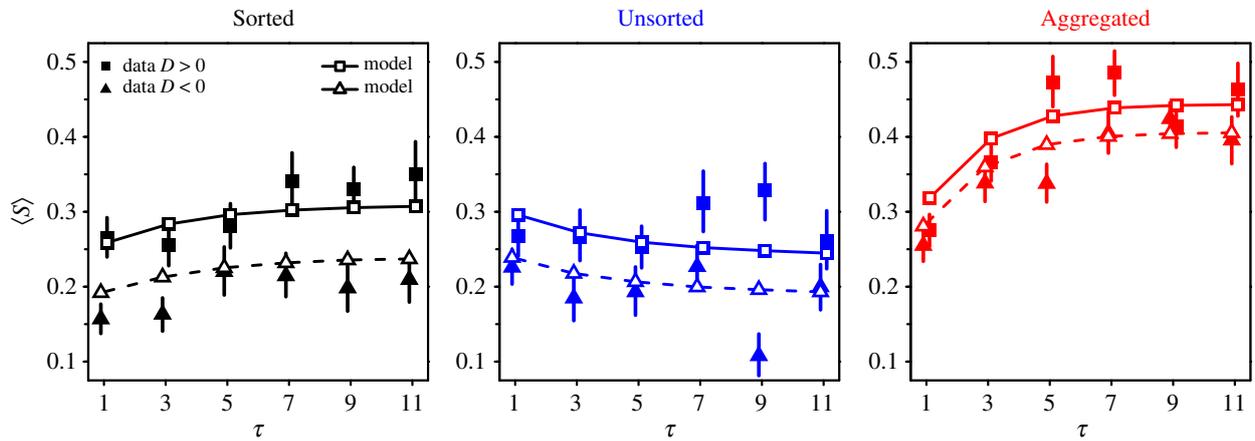

**Figure 8.** Average sensitivity to social influence $\langle S \rangle$ against the number of shared estimates $\tau$, in the Sorted (black), Unsorted (blue) and Aggregated (red) treatments, when the average social information $M$ is higher than the personal estimate $X_p$ ($D = M - X_p > 0$; squares) and when it is lower ($D < 0$; triangles). Subjects follow the social information more when $M$ is higher than $X_p$, than when it is lower. Filled symbols represent the data, while empty symbols (with solid lines for squares and dashed lines for triangles) are model simulations. Error bars were computed using a bootstrap procedure described in the Statistical methods, and roughly represent 1 s.e.

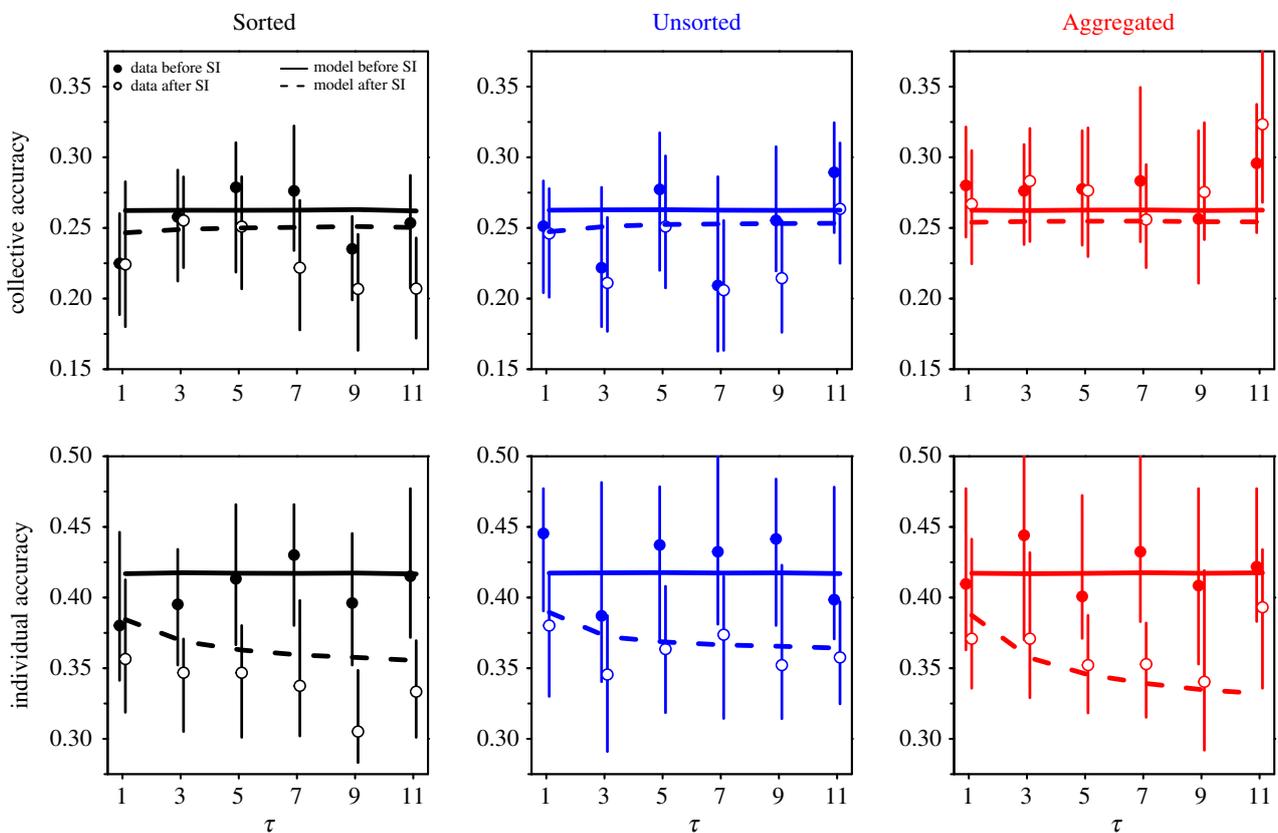

**Figure 9.** Collective and individual accuracy, against the number of shared estimates $\tau$, before (filled dots) and after (empty dots) social information sharing, in the Sorted (black), Unsorted (blue) and Aggregated (red) treatments. Values closer to 0 indicate better accuracy. Solid and dashed lines show model simulations before and after social information sharing, respectively. Error bars were computed using a bootstrap procedure described in the Statistical methods, and roughly represent 1 s.e.

information sharing, but without a shift of the median of the $X$, as was found in [20].

Figure 9 shows collective and individual accuracies against $\tau$ in each treatment before and after social information sharing. Collective accuracy improves, albeit marginally, in the Sorted and Unsorted treatments. This improvement is due to the higher use of social information when $D > 0$ than when $D < 0$ (figure 8). Because of a human tendency to underestimate quantities [20,31–33], weighing social information that is higher than one's personal estimate (i.e. $D > 0$) shifts individuals' estimates toward higher values after social information sharing, thus improving collective (and also individual) accuracy. In the Aggregated treatment, we find no such improvement in collective accuracy in the experimental data. However, the model predicts a slight improvement (due to $D_0 \lesssim 0$ and $\beta_+ \gtrsim \beta_-$). In all treatments, individual accuracy substantially improves after social information sharing. In the Aggregated treatment (which did not show any collective improvement), individual improvement is at par with both other treatments. This is due to the higher levels of social information use in the Aggregated treatment than in both other treatments (figure 7). Higher levels of social information use indeed result in a further narrowing of the distribution of estimates after social information exchange, thus bringing second estimates closer to the truth (see also figure 2). Moreover, the improvement in individual accuracy increases with the number of shared estimates $\tau$, and starts saturating at intermediate values of $\tau$. This increase and saturation follow the increase and saturation of $\langle S \rangle$ with $\tau$ (figure 7).

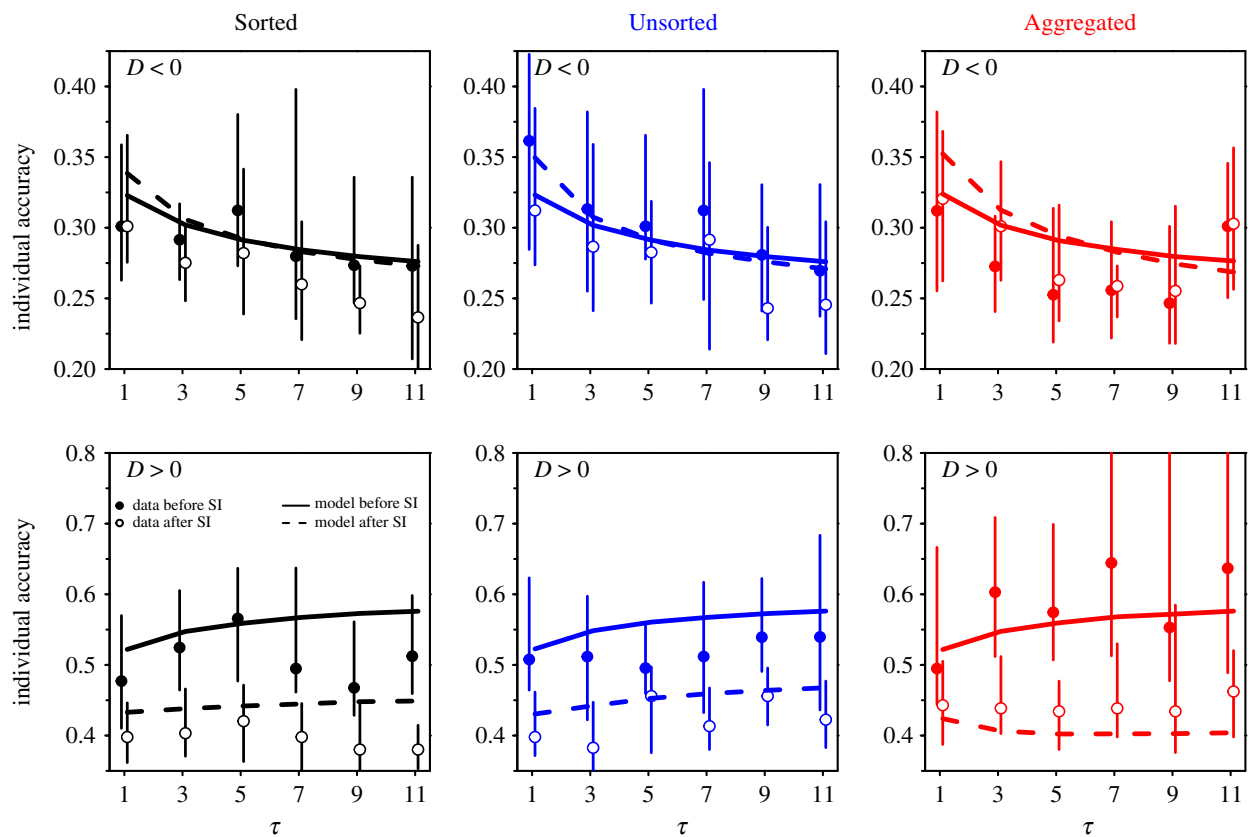

**Figure 10.** Individual accuracy against the number of shared estimates $\tau$, before (filled dots) and after (empty dots) social information sharing, in the Sorted (black), Unsorted (blue), and Aggregated (red) treatments. The data were separated into instances where subjects received social information of which the average ($M$) was lower than their personal estimate $X_p$ ($D = M - X_p < 0$) and instances of which the average was higher than their personal estimate ($D > 0$). Solid and dashed lines are model simulations before and after social information sharing, respectively. Individual accuracy hardly changes after social information sharing when $D < 0$, while it improves when $D > 0$. Error bars were computed using a bootstrap procedure described in the Statistical methods, and roughly represent 1 s.e.

Note that the dependence of collective and individual accuracy on $\tau$ in figure 9 is limited, in the data as well as in the model. The model, however, correctly predicts the range of values taken by individual and collective accuracy across all $\tau$, before and after social influence.

### 3.6. Impact of $D$ on estimation accuracy

Figure 10 shows individual accuracy when separating the data into instances where subjects received social information of which the average was lower than their personal estimate ($D < 0$), and where it was higher than their personal estimate ($D > 0$).

Interestingly, before social information sharing, individual accuracy was much higher (i.e. closer to 0) in the $D < 0$ case than in the $D > 0$ case. This is linked to the above-mentioned underestimation bias: when subjects receive social information that is lower than their personal estimate, their personal estimates are, on average, above the median of personal estimates and, therefore, relatively close to the truth. Vice versa, subjects who received social information higher than their personal estimate were, on average, below the median estimate and, therefore, farther from the truth.

When $D < 0$, individual accuracy remains unchanged (or improves marginally) in all treatments after social information sharing. When $D > 0$, however, individual accuracy improves substantially after social information sharing, and this effect is the main driver of the overall improvement in individual accuracy observed in figure 9. Electronic supplementary material, figure S6, shows the equivalent graphs for collective accuracy. Collective accuracy before social information sharing is also higher when $D < 0$ than when $D > 0$. Collective accuracy decays when $D < 0$, while it improves when $D > 0$, resulting in the overall mild improvement observed in figure 9. Despite the high level of noise (18 conditions plus dichotomy $D < 0$ versus $D > 0$), the model reproduces the experimental patterns fairly well.

### 3.7. Impact of $S$ on estimation accuracy

For each condition, figure 11 shows individual accuracy when separating the data into cases where sensitivity to social influence $S$ was lower than the median value of $S$ in that condition, and cases where $S$ was higher than the median value of $S$. In one instance (unsorted treatment at $\tau = 9$), the median of $S$ equalled 0, hampering an easy separation. In this case, we replaced the median by $S = 0.001$ (resulting in 50.8% (49.2%) of the data having an $S$ lower (higher) than this value). This value, and the experimental values of the median of $S$ in all other conditions, were used to separate the data in the model. In cases with relatively low values of $S$ (i.e. below the median), individual accuracy before social information sharing was higher than in cases with above-median values of $S$. This can be understood by considering that the lowest values of $S$ occur in situations in which the personal estimates are relatively close to the average social information (i.e. $D \sim 0$, see figure 5), which is also in general relatively close to the truth. However, because in those cases subjects did not (or barely) update their initial estimate, they did not improve in individual accuracy after social information sharing. By contrast, in cases with relatively high values of $S$ (i.e. above the median), subjects were able to improve in individual accuracy by using social information, their accuracy being at par with the below-median cases after social information sharing. In other words, information circulates from most to least informed individuals, thus evening accuracy across the whole group after social information sharing. These results are in line with previous findings [20,26], and are well predicted by the model. Electronic supplementary material, figure S7, shows similar, albeit less pronounced, results for collective accuracy.

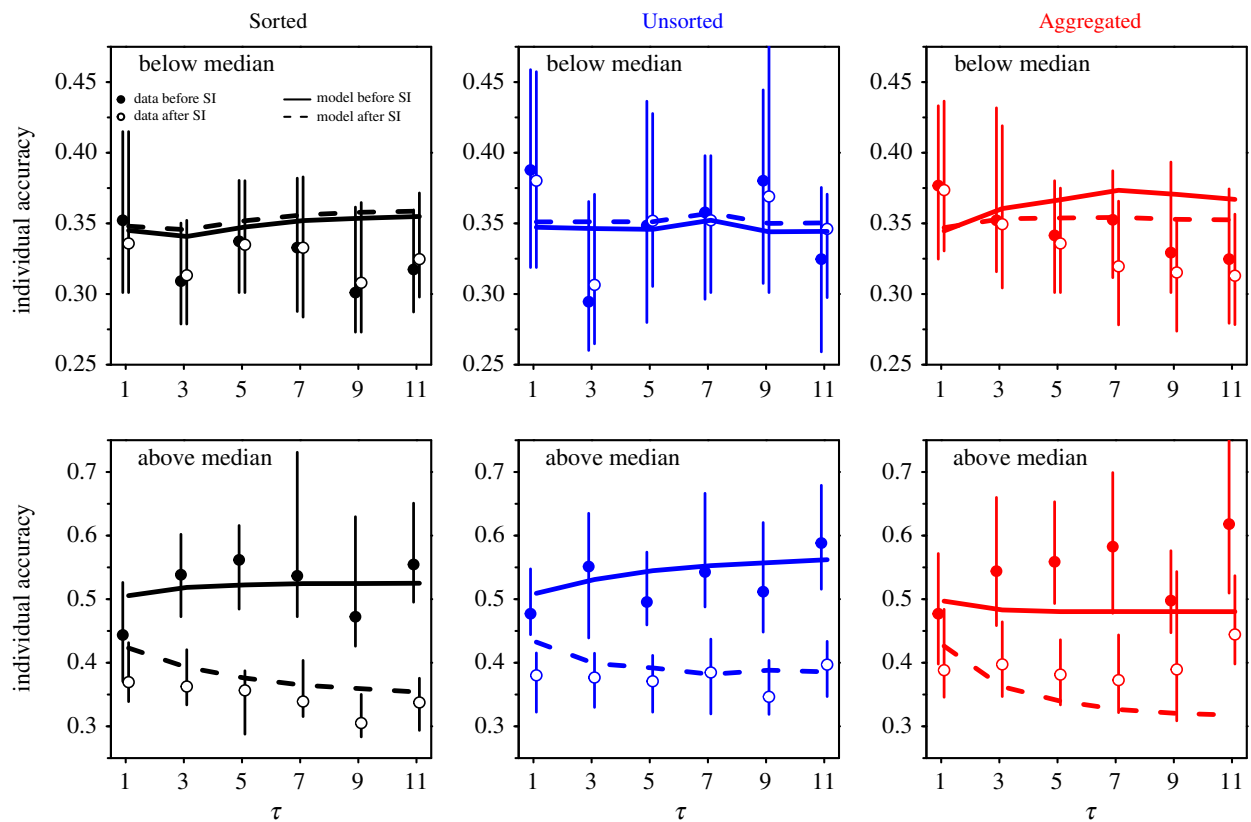

**Figure 11.** Individual accuracy against the number of shared estimates $\tau$, before (filled dots) and after (empty dots) social information sharing, in the Sorted (black), Unsorted (blue) and Aggregated (red) treatments. In each condition, the data were separated into sensitivities to social influence $S$ lower than the median value of $S$ in that condition, and sensitivities to social influence $S$ higher than the median value of $S$. Solid and dashed lines are model simulations before and after social information sharing. Cases with below-median values of $S$ (upper panels) had higher individual accuracy before social information sharing (i.e. values closer to 0) than cases with above-median values of $S$ (lower panels). The latter, however, used more social information and thereby improved more. Error bars were computed using a bootstrap procedure described in the Statistical methods, and roughly represent 1 s.e.

Note that in [20], the authors showed that there were individual differences in social information use across questions, with some individuals consistently disregarding the social information, and others consistently following it. They showed that the tendency to disregard social information correlated with confidence and concomitantly with accuracy before social information sharing. However, this average advantage of confident individuals disappeared after social information sharing, consistently with the present results.

### 3.8. Impact of group size on estimation accuracy

Finally, we use the model to generate predictions when sharing more estimates for larger group sizes. Electronic supplementary material, figure S8, shows how the number of estimates shared shapes improvements in collective and individual accuracy for groups of $N = 50$ individuals in all treatments.

In line with figure 9, improvements in individual accuracy are predicted to be substantially larger than improvements in collective accuracy. Moreover, the model predicts that individual accuracy improves more in the aggregated treatment than in both other treatments. This was, however, not observed in our empirical data with a group size of 12 (figure 9).

## 4. Discussion

The ever-increasing amount of information available online raises the question of whether exchanging aggregated forms of social information improves people's judgments more than making the complete information available. In the present work, we have compared the performance of groups in estimation tasks, when subjects received either all the available social information ($\tau$ estimates from other group members) or an aggregated version of it (the geometric mean of $\tau$ estimates from other group members).

We found that subjects follow social information substantially more when receiving the average estimate of other group members (Aggregated treatment), than when receiving a series of their individual estimates (Sorted and Unsorted treatments). Previous studies demonstrated that people are sensitive to the dispersion of estimates, and weigh social information more when the dispersion is low, a mechanism called the 'similarity effect' [22,30]. Possibly, observing the dispersion of the estimates in the Sorted and Unsorted treatments reduced the weight given to the social information as compared to the Aggregated treatment. In the Aggregated treatment, we further found that subjects' sensitivity to social influence increased sharply with the number of shared estimates $\tau$. This may reflect people's statistical intuition that the reliability of averages generally increases with the number of samples they are based on [34,35]. This tendency was observed to a much lesser extent in the Sorted treatment, and not at all in the Unsorted treatment.

Previous research demonstrated that subjects follow the social information more when it is farther from their initial estimate [20,26] (distance effect). Another way to interpret this is that subjects feel most confident in their personal estimate when the social information is close to it, reminiscent of a confirmation bias [36]. In [20,26], the social information consisted of the average of an *unknown* number of estimates from other group members. By contrast, subjects in our experiment received either a series of individual estimates from other group members (Sorted and Unsorted treatments), or were aware of the number of estimates used to compute the average (Aggregated treatment). This led to some important differences: in the Sorted and Aggregated treatments, subjects felt most confident in their initial estimate when that estimate was slightly above the geometric mean of the social information ($D_0 < 0$), rather than equal to it ($D_0 = 0$). This effect was weaker in the Aggregated treatment and absent in the Unsorted treatment. We also found an asymmetry in the distance effect in the Unsorted treatment, where subjects' sensitivity to social influence increased significantly faster when the geometric mean of the social information grew larger than their personal estimate, than when it grew lower ($\beta_+ > \beta_-$). This was not the case in the Sorted treatment, and barely in the Aggregated treatment. Both of these effects imply that subjects give more weight to the social information when it is higher than their initial estimate ($D > 0$) than when it is lower ($D < 0$). A similar asymmetry in social information use has been reported in [22,33]. Due to this asymmetry effect, subjects' second estimates shifted toward higher values, thus partly counteracting the underestimation bias [37–40], and improving collective and individual accuracy. Improvements in collective accuracy were mild in all conditions, whereas improvements in individual accuracy were much stronger. Note that improvements in individual accuracy in the Aggregated treatment were as high as in the other treatments despite the fact that improvements in collective accuracy were lowest in that treatment. This happened because improvements in individual accuracy are primarily driven by the level of social information use, which was substantially larger in the Aggregated treatment than in both other treatments. Indeed, a higher sensitivity to social influence leads to a narrowing of the distribution of estimates after social information sharing, and thus to second estimates getting (on average) closer to the truth.

We have built an agent-based model exploiting the empirical Laplace-like distributions of personal estimates and the distributions of sensitivities to social influence, as well as the distance effect. The model quantitatively reproduces the distributions of second estimates (figure 2), the dependence of the average sensitivity to social influence on the number of shared estimates (figures 7 and 8), and the improvements in collective and individual accuracy after social information sharing in all conditions (figure 9). The model also predicted correctly the difference in improvements in accuracy when the data were separated according to the social information received ($D < 0$ or $D > 0$; figure 10 and electronic supplementary material, figure S6) and according to people's sensitivity to social influence (figure 11 and electronic supplementary material, figure S7).

We used the model to generate predictions for larger group sizes and number of shared estimates. The model predicts that improvements in individual accuracy dominate those in collective accuracy, and are higher in the Aggregated treatment than in both other treatments. Moreover, improvements saturate with little marginal benefits beyond 20 shared estimates, even in the Aggregated treatment. These predictions are, however, based on a model built and calibrated for sharing 1–11 estimates. It is possible that changes occur when more estimates are shared, because people may use different strategies. We discuss these possibilities in more detail:

— In the Sorted treatment, one can reasonably expect people to keep on focusing on (and following) the central tendency of estimates when receiving more than 11 estimates. This is increasingly likely as more estimates are shared, since subjects would lack the time and cognitive capacity to assess (or even look at) all estimates. One may, however, expect that beyond a certain number of shared estimates, subjects would experience some sort of cognitive overload [41–43], and would not even be able to 'find' the central tendency of the social information. Since our model does not take into account cognitive overload, it may therefore not accurately describe how subjects integrate such large amounts of social information. If cognitive overload would arise at higher levels of $\tau$ than tested here, we would predict improvements in collective and individual accuracy to deteriorate compared to a cognitive overload free situation.
— In the Unsorted treatment, we can draw similar conclusions. We would, however, expect the cognitive saturation to happen faster as subjects are faced with unsorted estimates which are more difficult to process.
— In the Aggregated treatment, there is no cognitive limit to the number of estimates of which subjects receive the average. Our model may therefore be accurate at any group size. We, however, cannot exclude the possibility that the saturation 'breaks' at larger values of $\tau$, with social influenceability increasing again. Indeed, while it is reasonable not to expect much difference in subjects' behaviour when receiving the average of 10 or 15 estimates, it is less clear whether such stability would remain true between 10 and one million estimates.

In summary, we find that, in groups of 12 individuals, improvements in accuracy are similar when sharing all estimates or their geometric mean. We do, however, find that individuals use different strategies in these treatments to improve their estimates. Building a model based on our empirical results, and using this to generate predictions for larger group sizes, suggests that sharing averages would outperform sharing the full information in improving estimation accuracy at larger group sizes. Further research is, however, required to evaluate this prediction. Another avenue for future research would also be to investigate how other ways of displaying the social information fare compared to showing all estimates or their average. For instance, people may be able to better exploit the social information when observing graphical distributions, or other formats of aggregates. Finally, participants in our study were all undergraduate students, with homogeneous ages and levels of education. Future studies could investigate how sensitivity to social influence correlates with factors such as age, sex, culture, level of education or personality type. A partial attempt has been done in [20], where the authors showed that there was a negligible impact of sex on sensitivity to social influence, and that Japanese students tended to use slightly more social information than French students.

# 5. Statistical methods

## 5.1. Sensitivity to social influence S

### 5.1.1. Formal definition

We can write a subject's second estimate $X_s$ as the weighted arithmetic mean of their personal estimate $X_p$ and the social information $M$: $X_s = (1 − S) X_p + S M$. $S$ can thus be expressed as $S = (X_s − X_p)/(M − X_p)$. $S = 0$ implies that subjects keep their personal estimate ($X_s = X_p$), that is, they disregard the social information. $S = 1$ implies that their second estimate equals the geometric mean ($X_s = M$). In the Aggregated treatment, this implies that they perfectly follow the social information, and in the Sorted and Unsorted treatments, that they precisely adopt the central tendency of the social information (i.e. the geometric mean), which is highly unlikely to happen at $\tau > 1$.

### 5.1.2. Restrictions

Note that when $M \approx X_p$, $S$ can reach arbitrarily large values. For instance, if $X_p = 5$ and $M = 5.001$, then $X_s = 5.01$ gives $S = 10$ and $X_s = 5.1$ gives $S = 100$. Such large values of $S$ do not properly reflect the actual adjustment from $X_p$ to $X_s$, but grossly overestimate the level of social influence. Such outlying values can heavily affect measures based on $S$, in particular its average. To avoid this problem, we restricted $S$ to the interval [−1.05, 2.05] which excluded 4.32% of the data (these values were chosen rather than the more intuitive [−1, 2] for plotting purposes; figure 3). An additional 0.45% was excluded due to undefined values (which happened when $X_p = M = X_s$).

### 5.1.3. Fitting of the distribution

The following distribution was used to fit the distribution of $S$ (using the 'nls' function in R) and thus extract values of $P_g$, $m_g$ and $\sigma_g$ in all conditions:

$$f(S) = (1 − P_g) \, \delta(S) + P_g \, \varphi(S, m_g, \sigma_g) \quad (5.1)$$

and

$$\varphi(S, m_g, \sigma_g) = \frac{1}{\sqrt{2\pi}\sigma_g} \exp\left[-\frac{(S − m_g)^2}{2\sigma_g^2}\right], \quad (5.2)$$

where $P_g$ is fixed by equation (3.1), $\delta$ is the Dirac distribution centred on 0 and $\varphi$ is the Gaussian distribution of mean $m_g$ and standard deviation $\sigma_g$.

### 5.1.4. Fitting of the saturation

To fit the saturation observed in $P_g$, $m_g$ and $\sigma_g$ (figure 4), and extract their values in all conditions, we use exponential saturation functions where all three quantities saturate at the same rate $\varepsilon$:

$$Y(\tau) = Y_\infty − (1 − \varepsilon)^{\tau−1} (Y_\infty − Y_1), \quad (5.3)$$

where $Y = P_g$, $m_g$ or $\sigma_g$, $Y_\infty$ is the saturation value, $Y_1$ is the value of $Y$ at $\tau = 1$, and $0 < \varepsilon < 1$ is the saturation rate. The parameters $P_{g1}$, $P_{g\infty}$, $m_{g1}$, $m_{g\infty}$, $\sigma_{g1}$, $\sigma_{g\infty}$ and $\varepsilon$ were fitted simultaneously and their values are reported in table 1.

For consistency, $\alpha$ was fitted using the same functional form at the same rate $\varepsilon$ (figure 6).

## 5.2. Computation of the error bars

The error bars indicate the variability of our results depending on the $N_Q = 42$ questions presented to the subjects. We call $x_0$ the actual measurement of a quantity appearing in the figures by considering all $N_Q$ questions asked. We then generate the results of $N_{exp} = 1000$ new effective experiments. For each effective experiment indexed by $n = 1, ..., N_{exp}$, we randomly draw $N_Q' = N_Q$ questions among the $N_Q$ questions asked (so that some questions can appear several times, and others may not appear) and recompute the quantity of interest which now takes the value $x_n$. The upper error bar $b_+$ for $x_0$ is defined so that $C = 68.3\%$ (by analogy with the usual standard error for a normal distribution) of the $x_n$ greater than $x_0$ are between $x_0$ and $x_0 + b_+$. Similarly, the lower error bar $b_−$ is defined so that $C = 68.3\%$ of the $x_n$ lower than $x_0$ are between $x_0 − b_−$ and $x_0$. The introduction of these upper and lower confidence intervals is adapted to the case when the distribution of the $x_n$ is unknown and potentially not symmetric.

## 5.3. Fitting procedure used in figure 5

In our experiment, each combination of treatment and number of shared estimates contains 504 estimates. When binning data, one has to trade off the number of bins (thus displaying more detailed patterns) and the size of the bins (thus avoiding too much noise). In figure 4, the noise within each condition was relatively high when using a bin size below 1. However, bins of size 1 were hiding the details of the relationship between $\langle S \rangle$ and $D$, especially the location of the bottom of the cusp. To circumvent this problem, we use a procedure that is well adapted to such situations. First, remark that a specific binning leaves one free to choose on which values the bins are centred. For instance, a set of 5 bins centred on −2, −1, 0, 1, and 2 is as valid as a set of 5 bins centred on −2.5, −1.5, −0.5, 0.5 and 1.5, as the *same* data are used in both cases. Both sets of points produced are replicates of the same data, and therefore correct. But we now have 10 points instead of 5.

In each panel of figure 4, we used such a moving centre starting the first bin at −2, and the last one at +2, producing histograms (of bin size 1) in steps of 0.1 for the bin centre. This replicated the data nine times, thus having overall 10 replicates and 50 points, instead of 5. We then removed the values beyond $D = 2$, thus keeping 41 points ($D = −2$ to $D = 2$).

Next, we fitted these points using the following function:

$$\langle S \rangle_{\text{fit}} = \alpha + \beta_{\pm} |D - D_0|,$$

where $\alpha$, $\beta_-$, $\beta_+$ and $D_0$ are the fitting parameters. The nls function in R (used for the fits in the other graphs in the paper) had trouble converging because the absolute value function is not continuously differentiable.

Therefore, we wrote a program to directly perform the minimization of least squares. Let $\langle S \rangle_{\text{exp}}$ be the experimental values of $\langle S \rangle$ at each of the 41 values of $D$, and $Q = \sum_i \omega_i (\langle S_i \rangle_{\text{exp}} - \langle S_i \rangle_{\text{fit}})^2$ the weighted sum ($\omega$ is the fraction of data at each point), over all experimental points (indexed by $i$), of squared distances between $\langle S \rangle_{\text{exp}}$ and $\langle S \rangle_{\text{fit}}$. Equating to 0 the partial derivatives of $Q$ with respect to $\alpha$, $\beta_-$ and $\beta_+$ gave us analytical expressions of these parameters as functions of $D_0$. We then varied $D_0$ between −1.5 and 1.5 in steps of 0.01, and selected the parameter values with the lowest (weighted) sum of squared distances.

Ethics. All individuals signed an informed consent form prior to participating. The experiment was approved by the Institutional Review Board of the Max Planck Institute for Human Development (ARC 2018/08).
Data accessibility. The data supporting the findings of this study are available at figshare: https://doi.org/10.6084/m9.figshare.12529571.
Authors' contributions. B.J. and R.K. designed research and performed research; B.J. and C.S. analysed the data and designed the model; the three authors wrote the article.
Competing interests. We declare we have no competing interests.
Funding. No funding has been received for this article.
Acknowledgements. We are grateful to Felix Lappe for programming the experiment.